\documentstyle[prl,floats,aps,twocolumn,epsf]{revtex}
\tighten     

\begin{document}
\twocolumn[\hsize\textwidth\columnwidth\hsize\csname
@twocolumnfalse\endcsname   

\title{A superradiance resonance cavity outside rapidly rotating black holes}

\author{Nils Andersson$^{1}$ and Kostas Glampedakis$^{2}$}

\address{$^1$ Department of Mathematics, 
University of Southampton, Southampton SO17 1BJ, United Kingdom}
\address{$^2$ Department of Physics and Astronomy,
University of Cardiff, 
Cardiff CF2 3YB, United Kingdom}  

\maketitle

\begin{abstract}

We discuss the late-time behaviour of a dynamically
perturbed Kerr black hole. We present analytic results for 
near extreme Kerr black holes that show that 
the large number of virtually
undamped quasinormal
modes that exist for nonzero values of the azimuthal eigenvalue
$m$ combine in such 
a way that the field oscillates with an amplitude that decays
as $1/t$ at late times. This prediction is verified using
numerical time-evolutions of the Teukolsky equation. We argue that 
the observed behaviour may be relevant for 
astrophysical black holes, and that it can be 
understood in terms of the presence of a
``superradiance resonance cavity'' immediately outside the
black hole. 
\end{abstract}

\pacs{04.30.Nk}           

]  

{\em A brief background}. --- Our understanding of the
generic response of a black hole to dynamic perturbations is based on 
seminal work from 30 years ago. Exponentially damped
quasinormal-mode (QNM) ringing was first observed in numerical experiments
by Vishveshwara \cite{vishu}, and the subsequent 
late-time power-law fall-off (that all perturbative fields
decay as $t^{-2l-3}$ in the Schwarzschild geometry) 
was discovered by Price \cite{price}. A considerable body of 
work has since
established the importance of these two phenomena for black-hole
physics. 
We now know that most black-hole signals are  
dominated by the slowest damped QNMs, and many
reliable methods for investigating these modes have been 
developed \cite{novikov}. 
The nature of the late-time tail has also been studied in great 
detail. In particular it has been established that it is a generic
effect independent of the presence of an event horizon: The tail
arises from backscattering off of the weak gravitational potential
in far zone \cite{GPP}.  
However, the fact that our understanding has reached a 
mature level does not mean that  no problems remain
in this field.  A few years ago, 
the quasinormal modes had been calculated also for Kerr black holes
\cite{leaver}, 
but there were no actual calculations demonstrating the presence
of power-law tails. Neither were there any dynamical studies 
of rotating black holes.
Several recent developments have served to change this
situation and improve our understanding
of dynamical rotating black holes. Of particular relevance has been an effort
to develop a reliable framework for perturbative time-evolutions of Kerr 
black holes \cite{tcode}. There has also been recent efforts
to analytically approximate the late-time power-law tails for Kerr
black holes \cite{hod,leor}. Furthermore, 
numerical relativity is reaching a stage 
where fully nonlinear studies of spinning black holes are feasible. 

{\em Kerr black-hole spectroscopy}. --- With the likely advent of 
gravitational-wave astronomy
only a few  years away the onus is on theorists to provide
detailed predictions of the many scenarios that may lead to detectable
gravitational waves. In this context, the question whether
we can realistically hope to do ``black-hole spectroscopy'' by detecting
QNM signals and inverting them to infer
the black holes mass and angular momentum is highly relevant \cite{bhspec}.
For slowly rotating black holes this presents a serious challenge. 
Using standard results one can 
readily estimate that the effective gravitational-wave amplitude
for QNMs is (cf. similar estimates for pulsating stars \cite{akprl})
\begin{equation}
h_{\rm eff} 
\approx 4.2\times 10^{-24} \left( {\delta\over 10^{-6}}
 \right)^{1/2} \left(  { M \over M_\odot } \right)
\left( {15{\rm Mpc} \over r} \right) 
\end{equation}
where $\delta$ is the radiated energy as a fraction of the 
black-hole mass $M$. The frequency of 
the radiation depends on the black-hole mass as
$f \approx 12  ({M_\odot/ M}){\rm kHz }$.
Given these relations, and recalling the estimated 
sensitivity of the generation of detectors that is under construction,
the detection of QNM signals from slowly rotating solar-mass 
black holes seems rather unlikely. 
The situation 
will be rather different for low-frequency
signals from supramassive black holes in 
galactic nuclei and detection with LISA, the space-based interferometric
gravitational-wave antenna. Also,  there is recent
evidence that ``middle weight'' black holes, in the range 
$100-1000M_\odot$ may exist \cite{mediumbh}. For such black holes the
most important QNMs would radiate at frequencies where the new generation
of ground based detectors reach their peak sensitivity. If there are indeed
such black 
holes out there we may hope to take their fingerprints in the 
future.

It has been suggested that QNM signals from
rapidly rotating black holes would be easier 
to detect. This belief is based on  the
fact that some QNMs 
become very long lived as $a\to M$. In fact, mode calculations predict
the existence of an infinite set of essentially undamped modes
in the extreme Kerr limit \cite{leaver}. 
The available investigations into the detectability of QNM signals
have focused on the slow damping of these modes \cite{bhspec}. 
It has been shown that the decreased 
damping of the mode may increase the detectability
considerably. However, these results have to be 
interpreted with some caution. 
What has been shown is the (anticipated) effect that a slower damped mode
is easier to detect than a short-lived one, {\em provided that the modes
are excited to a comparable amplitude}. This is a rather subtle issue that
pertains to the question whether it is easier to excite a slowly damped
QNM than a short-lived one. Intuitively, one might expect
this not to be the case. 
In similar physical situations 
the build-up of energy in a long-lived resonant mode 
takes place on a time-scale similar to the 
eventual mode damping. Thus 
it ought to be very difficult to excite a QNM that has characteristic
damping several times longer than the dynamical 
timescale of the excitation process. 
This argument suggests that 
the amplitude of each long-lived mode ought to vanish in the limit 
$a\to M$ when the e-folding time of the mode increases dramatically
\cite{ferrari}.
In view of this it would seem rather dubious to conclude that the 
detectability of a QNM signal actually improves as $a\to M$. 
All may not be lost, however, because 
even if each individual QNM has an 
infinitesimal amplitude for rapidly spinning black holes 
a large number of modes approach the same limiting frequency
as $a\to M$. These modes may combine
to give a considerable signal \cite{sasaki}.

{\em A surprising analytic result}. ---
We want to assess the change in ``detectability'' of the QNMs as 
$a\to M$, i.e. as we approach the extreme Kerr black hole case.
As a suitable model problem, we  consider a massless 
scalar field. As is well known, the equation that governs such a field
 (which follows
immediately from $\Box \Phi=0$)
is similar to the master equation for both electromagnetic and gravitational
perturbations of a rotating black hole that was first derived by 
Teukolsky \cite{teuk}. 
In the following we briefly outline our calculation and 
discuss the main results. A more exhaustive discussion will be presented
elsewhere. We use standard Boyer-Lindquist coordinates,
and approach the QNM problem in the frequency domain 
(obtained via the integral transform used in \cite{andersson}). 
Furthermore, we use the symmetry of the problem to separate the
dependence on the azimuthal angle $\varphi$. In essence, we are
using a decomposition;
\begin{equation}
\Phi =  {e^{im\varphi} \over 2\pi }  \sum_{l=0}^\infty 
\int_{-\infty}^{+\infty} { R_{lm}(\omega, r) \over \sqrt{r^2 + a^2}} 
S_{lm}(\omega,\theta) e^{-i\omega t} d\omega  
\end{equation}
It should be noted that the rotation of the black hole
couples the various multipoles through the (frequency 
dependent) spheroidal angular
functions $S_{lm}$ \cite{teuk}.

In direct analogy with the Schwarzschild case \cite{andersson}
the initial value problem for the scalar field can be solved using a
Green's function
constructed from solutions to the homogeneous radial equation for 
$R_{lm}(\omega,r)$. One of the required solutions, that
satisfies the causal condition at the event horizon $r_+=M+\sqrt{M^2-a^2}$, has asymptotic behaviour
\begin{equation}
R_{lm}^{\rm in} \sim \left\{ \begin{array}{ll}
e^{-ikr_\ast} \quad \mbox{as } r\to r_+ \ , \\ A_{\rm
out} e^{i\omega r_\ast} + A_{\rm in} e^{-i\omega r_\ast} \quad
\mbox{as } r\to +\infty \ .
\end{array} \right.
\label{ingoing}\end{equation} Here
\begin{equation}
k = \omega - {ma \over 2Mr_+} = \omega - m \omega_+ \ ,
\end{equation}
where $\omega_+$ is the angular velocity of the event horizon, and
$r_\ast$ is the tortoise coordinate.
It is useful to recall that a monochromatic wave is superradiant 
if it has frequency in the range $0< \omega <  m \omega_+ $
\cite{teuk}.

A QNM is defined as a frequency $\omega_n$ 
at which $A_{\rm in}=0$. Assuming that 
$A_{\rm in} \approx (\omega - \omega_n) \alpha_n$ close
to $\omega = \omega_n$
we can deduce (via the residue theorem) 
that the contribution from each such mode to the 
evolution of the scalar field is
\begin{equation}
\Phi_n(t,r,\theta) =  {A_{\rm out} \over 2\omega_n  \alpha_n}
e^{-i\omega_n(t- r_\ast )} \sum_{l=0}^\infty
S_{lm}(\omega_n,\theta) {\cal I}_{lm}
\label{field}\end{equation}
where  ${\cal I}_{lm}(\omega_n,r)$ is a complicated expression that
depends on the details of the initial data (here assumed to have support
only far away from the black hole), cf. \cite{andersson}. 

Let us now focus on the case of nearly extreme Kerr black holes, 
i.e. on the case $a\approx M$. Then we can benefit from an approximation
due to Teukolsky and Press \cite{teuk}, that suggests that there
will exist an infinite set of QNMs that can be approximated by 
\cite{detweiler,sasaki}
\begin{equation}
\omega_n M \approx {m\over 2} - {1\over 4m}e^{(\theta - 2n\pi)/
2\delta} (\cos \varphi - i
\sin \varphi) 
\end{equation}
where $\delta$, $\theta$ and $\varphi$ are positive 
constants (not to be confused
with the coordinates), and $n$ is an integer labelling the modes. 
It is easy to see
that as $n\to \infty$ the modes become virtually undamped, 
and that they are located close to the upper limit of the superradiant
frequency interval. 
That such a set of long-lived QNMs will exist agrees with other 
mode-calculations \cite{detweiler,leaver}. 
Given the location of the QNMs we can extend 
the  calculation to deduce also the form of 
the asymptotic amplitudes $A_{\rm out}$ and $A_{\rm in}$ (or rather, 
the coefficient $\alpha_n$) for each $\omega_n$.
This enables us to approximate the contribution of each 
long-lived QNM to the field via (\ref{field}).  Doing this we find that
the longest lived modes have exponentially small amplitudes.
Thus we predict that the individual QNM will not in general be excited to 
a large amplitude, in agreement with the intuitive 
expectation. Expressing this result in terms of the effective amplitude
of a corresponding gravitational-wave QNM, we would have
\begin{equation}
h_{\rm eff} \sim \sqrt{ { \mbox{Re } \omega_n  \over \mbox{Im }
\omega_n } } {A_{\rm out}\over \alpha_n} \sim e^{-n\pi/2\delta}
\end{equation}
In other words, the assumption that the long-lived modes may be easier
to detect than (say) their short-lived counterparts for slowly rotating
black holes is cast in serious doubt. A recent, more detailed, 
calculation of the QNM excitation
coefficients  for $a\le M$ supports this conclusion.

This does not, however, mean that the long-lived QNMs are without relevance.
On the contrary, the fact that there is a large number of 
such modes has a very interesting consequence. After combining all
the long-lived modes we find 
\begin{equation}
\sum_{n=0}^\infty { A_{\rm out} \over \alpha_n} e^{-i\omega_n(t-
r_\ast)} \sim {e^{-im\omega_+ t} \over t} \quad \mbox{ as } t \to \infty
\label{tail}\end{equation}
This is an unexpected result: It suggests that, when summed, the
contribution from the slowly
damped QNMs of a near extreme Kerr black hole corresponds to a oscillating 
signal whose magnitude falls of with time
as a power-law. Furthermore, the decay of this signal is considerable
slower than the standard power-law tail. The decay of $1/t$ should be 
compared to the tail-results of, for example, Ori and Barack
\cite{leor} that suggests
that $\Phi \sim t^{-l-|m|-3-q}$
where $q=0$ for even $l+m$ and 1 for odd $l+m$ (derived only for non-extreme
black holes).
Hence, we predict that the oscillating QNM-tail 
will dominate the late-time behaviour of a perturbed 
near extreme Kerr black hole.

{\em Numerical confirmation}. --- 
Our analytic result is obviously surprising. However, 
in view of the many 
approximations involved in the derivation of (\ref{tail})  
considerable caution is warranted, 
and an alternative confirmation of the analytic prediction is desirable.
Fortunately, the recent effort to develop a framework for doing
perturbative time-evolutions for Kerr black holes \cite{tcode}
provides the means for testing our result. Hence, we have performed a 
set of evolutions (for various values of $m$)
using the same scalar field code that was used to study
superradiance  in a dynamical context \cite{tcode}.
As initial data we have chosen a generic Gaussian pulse originally
located far away from the black hole.

\begin{figure}[tbh]
\centerline{\epsfxsize=9cm \epsfbox{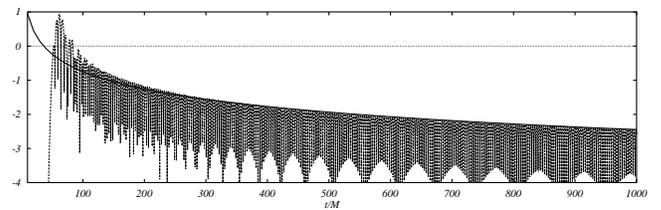}} \caption{A 
numerical evolution showing the late-time behaviour of a scalar
field in the geometry of a rapidly rotating black hole. We show
the field as viewed by an observer situated well away from the 
black hole for $a=M$. At late times the field falls off according to 
an oscillating
power-law with the amplitude decaying as $1/t$.  }
\label{fig1}\end{figure}

Our numerical evolution results can be succinctly summarised as
follows (further
 details will be discussed
elsewhere):

i) For extreme Kerr black holes ($a=M$) the numerical evolutions
show the predicted oscillating $1/t$ behaviour for all $m\neq 0$,
cf. Figure~\ref{fig1}.

ii) For $a<M$ we find a similar behaviour, i.e. that the
field is well approximated oscillations whose
amplitude decays as $t^{-\mu}$,
with $\mu$ rapidly increasing from 1 as $a$ departs from $M$, at late times.

iii) For axisymmetric perturbations ($m=0$) the numerical evolution
recovers the standard power-law tail. For our particular choice of
initial data (that contains the $l=0$ multipole) the tail
falls off as $t^{-3}$.  

Our interpretation of these results are: Firstly, the numerical 
evolutions confirm the analytic prediction for extreme Kerr
black holes, i.e. that the field
will oscillate with an amplitude that decays as $1/t$ at very late 
times. Secondly, and more important physically, the numerical data
suggests that the late-time behaviour is qualitatively similar
also in cases when $a$ is significantly smaller than $M$. The 
late-time behaviour was found to be consistent with an oscillating
tail in all cases
we have considered so far (essentially $a\ge 0.9M$). 
We have also verified that the observed 
late-time behaviour 
cannot be accounted for by a single slowly damped QNM when $a<M$. 
It is important
to emphasise that the result for $a<M$ was not predicted by the analytical
work (since the approximate modes we used are relevant only
for $a\approx M$). In other words, the numerical experiments indicate
that the effect could be of relevance for all
rapidly rotating black holes and may well dominate the standard 
power-law tail
for a large range of astrophysical black hole parameters.
Intuitively one would expect there to exist a critical value 
of the rotation parameter $a$ above which the new effect becomes
relevant. More detailed numerical work is needed to establish this, 
and investigate the role of the new effect further.

\begin{figure}[tbh]
\centerline{\epsfxsize=8cm \epsfbox{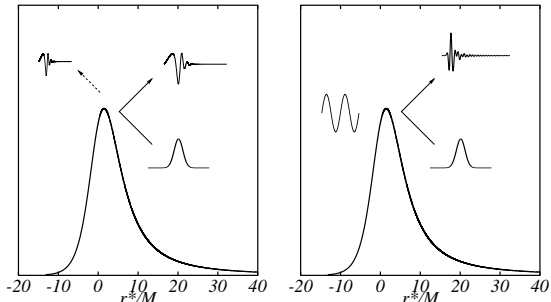}}
 \caption{Schematic explanation of the new phenomenon seen in 
the numerical evolutions of Kerr perturbations. The left panel
illustrates the standard scenario: An infalling pulse excites the QNMs that then
propagate to infinity and the horizon. At late times, backscattering due
to the curvature in the far-zone dominates and leads
to the familiar power-law tail behaviour. 
Right panel: Frequencies 
that lie in the superradiant regime 
experience a 
``potential peak'' in the region $[r_+,\infty]$. Hence,  there will be a 
superradiance resonance cavity 
outside the black hole. At late times, the waves 
leaking out of this cavity to infinity dominate the signal. }
\label{fig2}\end{figure}

{\em A physical interpretation}. --- 
Given both the analytic prediction for extreme Kerr black holes and the 
numerical evolution results for $a\le M$, an intriguing picture emerges.
The results seem to suggest the existence of a new phenomenon
in black-hole physics, with relevance at late times. We recall that 
the QNMs are typically interpreted, in analogy with scattering resonances in 
quantum physics, as originating from waves that 
are temporarily trapped close to the peak of the curvature potential 
(corresponding to 
the unstable photon orbit at $r=3M$ in the Schwarzschild spacetime), and that 
the late-time power-law tail arises because of backscattering off of the weak
potential in the far zone. Can the present results be interpreted in 
a similar intuitive vein? We think they can, and propose
the following explanation: Consider the fate of an essentially monochromatic 
wave that falls onto the black hole. Provided that the frequency is in the
interval $0< \omega < m\omega_+$ the wave will be superradiant. 
In effect, this means that a distant observer will see waves ``emerging
from the horizon'', cf. (\ref{ingoing}), even though a local 
observer sees the waves crossing the event horizon (at $r_+$) \cite{teuk}. 
This results in the
scattered wave being amplified. In addition to this, one can establish
that the effective potential has a peak outside the black hole
(which is not immediately obvious since the  potential is frequency 
dependent in the Kerr case) 
for a range of frequencies including the 
superradiant interval. Now, the combination of the causal boundary condition 
at the horizon effectively corresponding to waves 
``coming out of the black hole'' (according to a distant observer) 
and the presence of a potential
peak  leads to waves potentially being trapped in the region 
close to the horizon. In effect, there is 
a ``superradiance resonance cavity'' outside the 
black hole.  Again according to a distant observer, 
waves can only escape from this cavity by leakage 
through the potential barrier to infinity. Presumably it is this 
leakage that then
leads to the observed $1/t$ decay. Furthermore, it should be noted that
superradiant amplification is strongest for
frequencies close to $m\omega_+$. Thus  
superradiance  effects a form of parametric amplification on
waves in the cavity. At very late
times, the dominant oscillation frequency ought to be that which 
experiences the strongest amplification, i.e. $m\omega_+$. 
This is, of course, exactly  the result of our analytic calculation.   
The above argument is illustrated schematically in Figure~\ref{fig2}.

{\em Concluding remarks}. --- We have presented the results
of an investigation into the late-time behaviour of a
perturbed Kerr black hole. An analytic calculation for the near extreme
Kerr black hole case led to two important results.
Firstly, we deduced that even though 
some QNMs become very slowly
damped as $a\to M$ these modes will not 
be easier
to detect with a gravitational-wave detector. 
Secondly, we arrived at the
rather surprising prediction that the large number of virtually
undamped QNMs that exist for each value of $m\neq 0$ combine in such 
a way that the field oscillates with an amplitude that decays
as $1/t$ at late times. This decay is considerably 
slower than the standard power-law tail.
The analytic prediction was then verified using
numerical time-evolutions of the Teukolsky  equation.
These evolutions, performed for a larger range of the black-hole
rotation
parameter, suggest that the observed behaviour may well be present
also for astrophysical black holes (which we recall must have 
$a \le 0.998M$ \cite{kip}).
Finally, we proposed an intuitive explanation of the
observed phenomenon: That waves of certain frequencies
are effectively trapped in a 
``superradiance resonance cavity'' immediately outside the
black hole.  In conclusion, we find these results
tremendously exciting: They indicate the presence of a 
new phenomenon in black-hole physics that may well be of 
astrophysical relevance. 

We acknowledge helpful discussions with Amos Ori and Leor Barack. KG
thanks the State Scholarship Foundation of Greece for financial support.


\begin{references}

\bibitem{vishu}  C.V. Vishveshwara {\em Nature} {\bf 227} 936 (1969)

\bibitem{price} R.H. Price {\em Phys. Rev. D} {\bf 5} 2419 (1972); 
{\bf 5}  2439 (1972)  

\bibitem{novikov} Chapter 4 in ``Black-hole physics'' by V.P. Frolov and 
I.D. Novikov (Kluwer, Dordrecht 1998)

\bibitem{GPP} C. Gundlach, R.H. Price and J. Pullin {\em Phys. Rev. D} 
{\bf 49} 883 (1994);
E.S.C. Ching,  P.T. Leung, W.M. Suen and K. Young {\em Phys. Rev. D}
 {\bf 52}, 2118 (1995)
  
\bibitem{detweiler} S. Detweiler  
{\em Astrophys.\  J}.\  {\bf 225} 687 (1978)    

\bibitem{leaver}
E.W. Leaver  {\em Phys.\ Rev. D}  {\bf 34} 384 (1986)

\bibitem{tcode} W. Krivan, P. Laguna and P. Papadopoulos {\em Phys. Rev D}
{\bf 54} 4728 (1996);
 W. Krivan, P. Laguna, P. Papadopoulos and N. Andersson
 {\em Phys. Rev. D} {\bf 56} 3395 (1997); 
N. Andersson, P. Laguna and P. Papadopoulos {\em Phys. Rev. D} {\bf 58}
087503 (1998)   ; W. Krivan and R.H. Price {\em 
Phys. Rev. D} {\bf 58} 104003 (1998); {\em Phys. Rev. Lett.} {\bf 82} 1358 
(1999)    

\bibitem{hod} S. Hod. {\em Phys. Rev. D} {\bf 58} 104022 (1998) [also 
preprint gr-qc/9907096]

\bibitem{leor}  L. Barack and A. Ori {\em Phys. Rev. Lett.} {\bf 82}
4388 (1999) [also preprint gr-qc/9907085]; L. Barack preprint gr-qc/9908005

\bibitem{bhspec} F. Echeverria {\em Phys. Rev. D} {\bf 40} 3194 (1989);
L.S. Finn {\em Phys. Rev. D} {\bf 46} 5236 (1992); E.E. Flanagan and 
S. A. Hughes {\em Phys. Rev. D} {\bf 57} 4535 (1998)

\bibitem{akprl}  N. Andersson and K.D. Kokkotas  {\em Phys. Rev. Lett.}
{\bf 77} 4134 (1996)

\bibitem{mediumbh} A. Ptak and R. Griffiths {\em Ap. J. Lett.} {\bf 517}
85 (1999); E.J.M. Colbert and R.F. Mushotzky  {\em Ap. J.} {\bf 519}
89 (1999)

\bibitem{ferrari} V. Ferrari and B. Mashhoon
 {\em Phys. Rev. D } {\bf 30} 295 (1994)  

\bibitem{sasaki}
M. Sasaki and T. Nakamura {\em Gen. Rel. Grav.} {\bf 22} 1351 (1990)  

\bibitem{teuk} S.A. Teukolsky {\em Phys.\  Rev.\  Lett}. {\bf 29}
1114 (1972); S.A. Teukolsky and W.H.  Press {\em Astrophys.\  J.}
{\bf 193} 443 (1974) 
        
\bibitem{andersson} N. Andersson 
 {\em Phys. Rev. D} {\bf 55} 468 (1997)       

\bibitem{kip} K.S. Thorne {\em Astrophys. J.} {\bf 191} 507 (1974)

\end{references}
\end{document}